\documentclass[12pt]{article}
\usepackage{amsmath,amsfonts}
\usepackage[numbers]{natbib}
\usepackage{hyperref}
\usepackage{graphicx}
\usepackage[margin=0.75in]{geometry}

\begin{document}
\date{\today}
\title{{\Large Origin Gaps and the Eternal Sunshine of the Second-Order Pendulum}}
\author{Simon DeDeo\footnote{Department of Social and Decision Sciences, Carnegie Mellon University \& the Santa Fe Institute. {\tt sdedeo@andrew.cmu.edu}; \url{http://santafe.edu/\~ simon}. FQXI Prize Essay 2017.}}
\maketitle

%\noindent {\it Wandering Towards a Goal: How can mindless mathematical laws give rise to aims and intention?}
\begin{abstract}
\noindent The rich experiences of an intentional, goal-oriented life emerge, in an unpredictable fashion, from the basic laws of physics. Here I argue that this unpredictability is no mirage: there are true gaps between life and non-life, mind and mindlessness, and even between functional societies and groups of Hobbesian individuals. These gaps, I suggest, emerge from the mathematics of self-reference, and the logical barriers to prediction that self-referring systems present. Still, a mathematical truth does not imply a physical one: the universe need not have made self-reference possible. It did, and the question then is how. In the second half of this essay, I show how a basic move in physics, known as renormalization, transforms the ``forgetful'' second-order equations of fundamental physics into a rich, self-referential world that makes possible the major transitions we care so much about. While the universe runs in assembly code, the coarse-grained version runs in LISP, and it is from that the world of aim and intention grows.
\end{abstract}

\noindent 
\begin{verse}
{\it How happy is the blameless vestal's lot!  \\
The world forgetting, by the world forgot.  \\
Eternal sunshine of the spotless mind! \\
Each pray'r accepted, and each wish resign'd} \\
--- Alexander Pope, ``Eloisa to Abelard''\footnote{\url{http://bit.ly/2BQganC}}
\end{verse}

\vspace{0.5cm}
\noindent The world we see, and the worlds we infer from the laws of physics, seem completely distinct. At the blackboard, I infer that a thin skein of gas will coalesce into objects such as stars and galaxies. With a few more assumptions I predict the range of masses that those stars should have, beginning from an account of initial quantum fluctuations. Today, it's considered a reasonable research goal to reduce even that story, of the wrinkles in spacetime that seeded Andromeda, to the first principles of basic physics: Hawking radiation at a horizon, the quantum statistics of a multiverse.

If, however, I try to infer the existence of the blackboard itself, and the existence of people who write on it and themselves infer, I am stuck. I find myself unable to predict the spectrum of desires and goals that evolution can produce, let alone the ones that arise, apparently spontaneously, from the depths of my own mind. The utter failure of otherwise reliable tools to generalize to this new domain is one that many scientists experience when they cross between fields. Not just scientists: as Sherry Turkle pointed out, even young children experience it, when confronted by electronic toys. There is something about the experience of life (or life's substrate, computation) that goes beyond purely physical mechanisms they're used to seeing in other toys. A child faced with an apparently living machine looks in the battery compartment to see what powers it~\cite{turkle2011life}. Whether it is felt by an adult scientist at the blackboard, or a child with a toy robot, it is at heart an experience of the gap between the purposeful world of human life and the aimless one of stars. Our tools can not make the leap.

Our tools do, of course, work if we are allowed to assume the existence of meaning-making beings to begin with. Fluid dynamics can describe the flow of traffic through my city, while variants on the Ising model allows me to predict the racial segregation I see as I pass through it, and further generalizations get us off on the right foot for thinking about how my messily-wired brain might learn and remember and experience at all. 

Yet no matter how well we do once meaning-making beings are taken as a given, we stumble when we are asked to predict their very being at all. It is \emph{this} gap, the inability to leap from one side to the other, that begs explanation, and I refer to it as the Origin Gap because it is familiar to those working in the ``origin'' fields: the origin of society, the origin of consciousness and meaning, the origin of life. It is the gap that gives those fields a very different flavor from the sciences of their mature subjects. Origin of society looks very different from social science and anthropology; origin of consciousness looks very different from psychology; origin of life looks very different from biology.

%The gap is real, but the fields that try to bridge it are far from hopeless endeavours. While scientists struggle with them, it is worth examining the gap itself more closely and from an abstract point of view. Doing so allows us to get clear about exactly what the origin fields have as their special domain and special challenges. If we understand what exactly is hard about the origin fields, researchers interested in these deep questions will be more likely to actually spend their time on origin stories---rather than reinventing, in an inefficient and {\it ad hoc} manner, bits and pieces of biochemistry, neuroscience, or sociology that biochemists, neuroscientists, and sociologists could do better.

The gap, I claim, is understandable, even (one might say) predictable. In this essay, I'll first show that the existence of the gap is the consequence of a basic pair of facts in the theory of computation. Second, that particular aspects of the laws of physics make it very likely that in the evolution of the universe, such gap will naturally appear. Taken together, these facts explain how ``mindless'' laws lead to the emergence of new realms of intentional behavior. At the heart of this essay's explanation of the gap will be that the kind of intention, aim, and meaning we really care about also has the capacity to refer to itself. 

\section{The Mathematics of the Gap}

From the mathematical point of view, the origin gap begins with the fact that
\begin{enumerate}
\item it is easy to describe everything. 
\item it is much harder to describe one thing.
\end{enumerate}
This has a counterintuitive feel to it. We began, both as individuals and as a species, by describing particular things (that big mountain, this frozen river, that tall woman, this cold morning). It therefore feels as if this task must be easier than the more elaborate habits of generalization, abstraction, the tools of set theory, category theory...

Yet when we make this leap, we forget how many millions of years of evolution went into teaching us how to produce these descriptions. What it means to be one thing rather than two, the identification of useful boundaries or persistent patterns, what it means for an argument to be valid: each is a question subject to endless debate. We see this in the history of philosophy, but a more contemporary example comes from the history of Artificial Intelligence (AI). AI gave a name to the feeling that rules of description could never be exhaustively specified. They called it the ``frame problem'', and advanced societies across the globe dumped literally billions of dollars into solving it. Until, that is, they discovered that the quickest way to solve the problem of describing something was to avoid specifying the rules at all.%\footnote{Not quite the real history. First, the frame problem was used to justify cutting off funding for AI altogether, at least in the United Kingdom---the so-called {\it AI Winter}. Only later, did people realize that people could solve the frame problem without direct attack.} 

Rather than define in computer code a beautiful sunset, or a valid argument, researchers now build learning machines that watch and copy human response. Don't describe a cat to a computer; have it learn what a cat is from the pictures we take to celebrate them on the internet. In this way, the code can rely on the accumulated wisdom of evolution. Which is only natural, since (of course) a computer is build by evolved creatures to serve their needs. 

In as much as our lives are dominated by artificial intelligence we have, for now,  given up on describing things. But it turns out that to describe {\it everything}, by contrast, is simple. It only needs to occur to you to do something so trivial as to try. Borges did so in his short story {\it The Library of Babel}, where he imagined a series of interconnected hexagonal rooms, walled by shelves and stacked with books, and each book containing the letters of the alphabet, spaces, commas, and periods in different orders.

How much is contained in everything! Of course, in Borges' library there are an overwhelming number of nonsensical books, cats typing on keyboards, but also (again, of course) the complete works of Shakespeare, as well as every variation on those works, and every possible edition with typographical errors, and (as Borges might have gone on) the plays that Shakespeare might have written were he really Francis Bacon, or Elizabeth the First, or an alien from Mars, as well as all the incorrect extrapolations of those conjectures, and so on to the limit of one's imagination, and (then) beyond.

Imagine that we have instant access to the text of any book. It's simple to find all the books that include the word ``Shakespeare'': just send your robot out to search book by book and return the ones that contain that string of letters. Of course, it will also recover nonsense books, books full of jumbled letters that happen, once in awhile, to spell the name: ``...casa,cWas,,,qwh g Shakespeare acqq CO...''

Here's a harder problem: how to locate the books on Shakespeare that make sense? Give instructions to the robot to gather them together. Or, imagine the layout of the Borges library as a wireframe image on your computer screen, and the rules of shelving to hand. Outline, or click with a mouse, the shelves to pull.

Under some very basic assumptions (which we'll address below), the strange thing about this more complicated query that it can imagined, but not actually made. Like the idea of squaring the circle, of producing using straight-edge and compass a square whose area is equal to a given circle, it seems that it should be possible. And yet it is not: the shape your mouse carves out, although imaginable in each fragment---``this book, not that''---is an impossible shape, a shape impossible to define and therefore to draw. In its infinitely detailed structure, it is at each scale completely unrelated to the scale above.

Computer scientists usually introduce these shapes in a very different fashion: by describing things that are capable of self-reference, and then by showing how questions about these self-referring things, though well-phrased, have no answers. Consider, for example, a game two mathematicians might play: name the number. ``The smallest prime greater than twelve", for example, names the number thirteen. ``Two to the power of fifty" names a much larger number, something just a bit bigger than 100 trillion. There are better and worse ways to name something (``one plus one plus one plus...'' is a poor way to start naming a number larger than ten thousand, say), and you might imagine mathematicians competing to name something in the most efficient, the shortest, way. 

A classic example of how the game goes wrong was provided, appropriately enough given our introduction of the Borgesian Babel Library, by an Oxford librarian, G.G. Berry, who asks us to consider the following sentence:
\begin{equation*}
\textrm{``The smallest number that can not be named in less than a thousand words."}
\end{equation*}
Such a sentence has a twisty logic to it: whatever it names, it certainly names in less than a thousand words. And yet whatever it purports to name must be something that actually requires the far larger sentence. The resolution of a paradox like this is not to reject the sentence, but to rule out the possibility of the efficient mathematician, a person (or machine) that finds the shortest description of any number. The problem that Berry's paradox reveals is that problem with certain kinds of systems that can refer to themselves: Berry's paradoxical sentence refers (implicitly) to the very practice it enacts, that of finding short descriptions. It's an easy matter for the impossibilities implied by self-reference in the Berry case to lead to the problems of locating books on Shakespeare in the Borgesian library. 

The existence of such shapes (or the non-existence of the rules of their construction) seems counter-intuitive at first, because it is the nature of human beings to ask for things that are possible. ``Bring me all the sugar in the kitchen"; ``Find me all the students in the engineering department". We are not used to asking questions that have no answer.

Yet for it to happen all we need is that any description of what it means to be a sensical book on Shakespeare requires more than just pattern-matching (\emph{e.g.}, the presence or absence of the word ``Shakespeare''). Impossible questions emerge when they become about pattern-processing, pattern manipulation, pattern computation. Something sophisticated enough, in particular to allow us to have something operate on a description of itself.

This is, of course, exactly what takes place. If we read a book on Shakespeare we do more than count words and match them to lists. We think about those words, the combinations they fall in, and what one combination means for another. We reason about a passage, follow its arguments and conjecture counterarguments. And when we give ourselves, or a machine, that power, we become fundamentally limited in the questions we can ask and answer about what we, or it, is going to do. It becomes impossible, even, to draw outlines around the behaviors we do, or do not, expect. In contrast to the condensation of gas into stars, we can not derive, ahead of time, the space of books that scholars will write about Shakespeare.

This is why the origin problems are hard. The things whose origins most intrigue us are also the points at which systems gain new powers of self-reference. And these moments lead to new categories, new phenomena, that we can literally not predict ahead of time. Once we have an example, we can ask questions about it, do science on it, just as we can take any particular volume from the Borgesian library and read it. But to begin with the space of all possible things that can happen, and then to draw the outlines of what we expect to see on the basis of a self-referential process, is something else altogether.

I'm hardly the first to draw attention to the importance of self-reference for the problems of life. Sara Walker and Paul Davies have pointed to the self-referential features at the heart of the origin of life problem~\cite{walker2013algorithmic}. Stuart Kauffman puts self-reference at the heart of both biological and social evolution, and in places conjectures explicitly G{\"o}delian arguments~\cite{kauffman2016humanity}. My own work, and that of my collaborators, on social behavior suggests that social feedback, the most primitive form of self-reference and something we see in the birds just as much as the primates~\cite{hobson2015social}, is at the origin of major transitions in political order~\cite{major}.

The gap between physics and the meaningful experiences we associate with life thus turns out to have an unexpectedly mathematical feel. The emergence of meaningful experiences is associated with the emerge of new forms of self-reference, but questions about the basic properties of self-referential systems are (on pain of logical inconsistency) impossible to answer in the complete and general fashion we expect from derivations in the physical sciences. 

Asked to sketch out the consequences of a new self-referential phenomenon---say, an organic polymer than can refer to, modify, and reproduce itself---we stumble, because the very question is unanswerable. Given a particular example (the replication machinery of the bacteria {\it E. coli}) we can do a great deal of science. But to delineate all the life this makes possible is equivalent to picking books of Borges' library.

Before moving to the next part of this essay's argument, the physics of self-reference, it's worth pointing to the leap that's implicitly being made here. The Earth, and everything on it, is finite in nature: only so many things will ever happen. If the holographic principle is true, we may even be able to compute the total entropy contained within the boundary of our planet's world line. This enables purists to object to the arguments I've made here, because G{\"o}delian impossibility theorems usually require an infinity somewhere. Explicitly, the things that self-reference makes impossible are those that are required to apply to everything in the domain in question: every number in the set of integers, every program that could be written and how it behaves on every set of inputs. All the numbers involved (the size of the Borgesian library, whose books are of limited size; indeed, the number of behaviorally-distinguishable possible configurations of the human mind) are not infinite, but rather simply very, very, very large. This means it is possible to tell your assistant what to do: you could, for example, go out yourself, read all the books shorter than a certain length, and give him a list. Irritating as these list-based solutions are, it's rather hard to rule them out. They're clearly unsatisfactory, because they somehow presume the answers are already to hand, a little like giving someone a grammar of the English language that simply lists all sentences shorter than ten thousand words. We want something that summarizes, compresses, or otherwise gives us rule-based insight. 

When infinities go away, however, it can become possible to approximate the things we want without falling victim to paradox. We often want to talk about ``shortest descriptions'', for example, even when their kind of twisty self-reference puts us in the cross-hairs of Berry's paradox. In a 2014 paper Scott Aaronson, Sean Carroll, and Lauren Ouellette squeezed around it by using the file compression program {\tt gzip}~\cite{aaronson2014quantifying}! It's a clever idea, and (in my opinion) an excellent way to probe the problem they have at hand, but it's not going to work in all situations: I shouldn't try to judge the complexity of a student's reasoning by looking at the filesize of a {\tt gzip}ped version of her text. While we have heuristics and good ideas in some situations, we don't yet have a good handle on how an impossible problem ``degrades'' into a solvable one more generally.  It's likely that the theory of computational complexity will play a role (see Ref.~\cite{aaronson2013philosophers} for a philosophical overview).

Noam Chomsky confronted this problem head on in \emph{Aspects of the Theory of Syntax}~\cite{chomsky2014aspects}, where he distinguished performance (what we say) and competence (grasping the rules of what we say), and introduced to linguistics the idea that good rules, the kind of rules we want, are ``generative''. Something like Chomsky's competence-performance distinction, and insistence on the creation of generalizable rules rather than the creation of endless descriptive categories, is part of the story.

\section{The Physics of the Gap}

It is one thing to ascribe the gap to the emergence of new systems of self-reference. But why should self-reference come into being at all? At the heart of self-reference is the existence of a memory device, and something that can navigate it in a ``sufficiently sophisticated'' fashion. Smith and Sz\'{a}thmary's famous 1997 piece~\cite{smith1997major}, on the major transitions in the biological record, recognizes this implicitly, placing the discovery of new information processing and recording mechanisms at the center of each transition. Social scientists~\cite{fukuyama2011origins}, scholars of ``deep history''~\cite{smail2007deep}, and cognitive scientists~\cite{bellah2012axial_merlin} each draw attention to new institutions, like cities, or new cultural practices, such as writing or social hierarchy, or even new abilities from physiology itself, such as genes for speech and syntactic processing, that enhance the ways in which we can remember and transform what we remember. Major leaps occur when something previously forgettable, lost to noise, finds a means to be recorded, translated into a referential form, processed and combined with others. When social debts become stories told around a campfire---or transform into money and markets~\cite{graeber2014debt}---we see not just an augmentation of life as it was known, but the unpredictable creation of entirely new forms of being.

Each of these moments is a shift in the nature of the world, and a clear topic of scientific study. Whether we study the details of its emergence, or the patterns it displays that generalize beyond its historical context, any one of them is the task of a lifetime. But what makes memory, and self-reference, possible at all?

Strangely enough, it's not baked in to the fundamental laws of physics, a fact that was driven home to me early in my career, at the University of Chicago, when I worked with Dimitrios Psaltis, and Alan Cooney, physicists at the University of Arizona. We were puzzling over a strange class of models in fundamental physics. Despite their mathematical coherence, they were, at heart, unstable: any universe that obeyed their laws would sooner or later explode, everywhere and instantaneously, into fountains of energy with no apparent end, as if slipping off the top of a hill that had no bottom.

What made them unstable was how they handled time. In the physics you encounter in high-school it's crucial that Newton's laws of motion talk about the relationship between force and acceleration: $F=ma$, force is mass times acceleration, or perhaps more easily, $a=F/m$, the acceleration you experience is the force applied to you, divided by your mass. Acceleration is connected to the passage of time; it's how fast your velocity is changing, or, more formally, the ``second derivative of position with time''. Newton's laws then connect forces you might experience to a phenomenon we call gravity: objects create a gravitational field, and at each point that field subjects objects to a certain amount of force. Other laws talk about other sources of force: electrical, or magnetic, for example. All connect back to acceleration, the change of velocity with time.

This is all awesome and highly addictive to talk about if you have a certain bent of mind, but one of the basic facts about these laws is that you never see anything with more than two derivatives in the fundamental equations. When you write them down, you only ever talk about (1) a basic set of quantities, say, position, gravitational field, etc.; (2) how these quantities change with time; and (3) sometimes, how these changes in time change with time. If you have a theory where higher derivatives enter in, where you talk about changes in changes in changes, then the theory becomes unstable in some really uncomfortable ways, leading to things like spontaneous infinite accelerations which you never observe (or really could imagine observing) and that would really ruin your day if you did. This has been known since the 1850s, when the Russian physicist Mikhail Vasilevich Ostrogradsky published what is now called ``Ostrogradsky's Theorem'' in the journal of the Academy of St. Petersburg. At the end of this essay I provide an afterword that gives the underlying physical intuition for why this is true, through an economic analogy to a shift in marginal costs. A technical introduction can be found in the account by physicist R.P. Woodard~\cite{woodard2015theorem}, while Dimitrios, Alan, and I were working on how to ``cure'' these instabilities in certain limited regimes; you can find our answers (and further references) in a series of papers we wrote together~\cite{dedeo2008stable,cooney2009gravity}.

Ostrogradsky's theorem sounded just fine to me, until I remembered something from my high school physics teacher. The change in acceleration, the \emph{third} time derivative of position, has its own name---``jerk''. Jerk is what you experience when an elevator starts up. When it's moving at a constant velocity, you feel nothing. When it's accelerating, you feel heavier (if you're going up), or lighter (if you're going down). But when it switches from not accelerating to accelerating, or vice versa, you experience a sudden change in your weight. You're experiencing the elevator jerking you up, or the pit of your stomach dropping out when it descends.

The fact that I experience jerk is very strange. Am I not a creature that lives in the physical world? Am I not forced to obey the laws of physics? And don't I know, from a bit of mathematics, that the laws of physics only deal with quantities with two time derivatives or fewer, or risk being violently unstable if they don't? But if all that's true, how can jerk, a third-order quantity, play any causal role in my life, such as causing me to say ``oof'', or making me feel queasy, when the elevator moves? How can my psychological laws obey equations that are ruled out as physical laws?

I remember a spooky feeling when I put this argument together, and for a brief moment wondering if this proved the existence of a separate set of psychological laws beyond or parallel to physics. The answer turns out to be a bit simpler, if no less intriguing. The instabilities that emerge for theories with higher-order derivatives are real, and barriers to them being basic laws of the universe are real as well. But there's nothing that prevents them holding for a while, in limited ways, so that the instabilities don't have time to emerge.

And that's the reason I can feel the jerk. I have a brain that senses acceleration. It's possible for that sense to rely directly on fundamental laws (it doesn't, actually, but it could). But in order to report the sensation of jerk to my higher-order reason, my brain has to go beyond fundamental physics. It has to use memory to store one sensation at one time and compare it, through some wetware neural comparison device, to a sensation at a later time. Similarly, I can measure the acceleration that my car undergoes by hanging a pendulum from the ceiling and seeing where it points, leveraging a little bit of fundamental physics. But to measure jerk, I have to videotape the pendulum, and compare its location at two different times. There's no ``jerk pendulum'' I can build that relies directly on the basic laws of physics that apply everywhere and for all time. The fundamental laws are forgetful, the ``blameless vestals'' of the Alexander Pope quotation that begins this essay.

It's strange to think that a visceral and immediate feeling, like the drop you feel in the pit of your stomach when the elevator descends, is an experience filtered through a skein of memories. These memories present what is actually a processed and interpreted feature of the world as if it were a brute physical fact. Yet it so turns out that some things, like ``force'', are truly fundamental constituents of our universe, while others, like ``jerk'', are derived and emergent.

Jerk gets into the physical world through memory, but it's hardly the most impressive feat of memory we do. A man descending a New York City skyscraper is in the presence of far greater feats of memory and processing than just what travels down his vagus nerve. Yet jerk also gives us a clue to how those far more sophisticated memories might have gotten going. The experience of jerk is an atavism of a far more primal event, one that began well before there were brains to feel it.

This is because, while (to the best of our knowledge) higher order ``memory terms'' like jerk are forbidden from playing a role in fundamental laws, they do emerge in an unexpected fashion. We rarely perceive the world at its finest grain, in all of its fluctuating detail, at the assembly code level, you might say as a computer programmer. We see, instead, averages: not everything that happens in a single patch, but a coarse-grained summary of it, a blurring of details as if the lens was smeared with vaseline. To give a full account of the role of that averaging, or coarse-graining, in the physical sciences would take us very far afield, but also (many now believe) into some of the best mysteries we have to hand, including an explanation of the second law of thermodynamics and the decoherence, or collapse, of the quantum-mechanical wavefunction.

Here we care about coarse-graining because, by averaging together nearby points, it introduces the possibility of inducing physical laws that (in contrast to their forgetful fundamental cousins) do have memory. When we smooth out the world, when we average out some of the small-scale bumps and fluctuations, we produce a new description of it. The laws that govern those coarse-grained descriptions, in contrast to the ones that applied at the shorter distances and for the finer details, can have memories, can include higher derivatives. They may, in certain cases, be unstable, but this is no longer an existential threat: it just means that, occasionally, the coarse-grained description will fail. The fine-grained details will emerge with a vengeance, ruining the predictive power of the theory. You'll be reminded of the limits of your knowledge, but the universe will not catch fire.

The technical term that physicists use for this is renormalization. Physicists use it for all sorts of problems, and call the theories that emerge for coarse-grained systems ``effective theories''. My colleagues and I have thought about them for a long time, as both a fact of life for deriving one scale from another (social behavior, say, from individual cognition), and a metaphor to help explain why biology differs so much from biochemistry, and why averaging-out might not just be a good idea, but might make new forms of society possible~\cite{dedeo2011effective,flack2012multiple,hobson2015social,major,Flack20160338}.

If you're a computer scientist, you might say that while the universe runs in assembly code, the coarse-grained version runs in LISP. Here, coarse graining gives the possibility of memory and---with some interesting dynamics for how those memories inter-relate---the self-reference that makes certain features of the future logically unpredictable based on what came before. The memories we have now, biochemical, electronic, on pen and paper and in the cloud, are far more complex than the ones than appear in a physicist's coarse-graining prescription. 

You get a great deal from the averaging-out a cell wall allows you to do, another boost from the ways in which neurons average out the data from your eye, and another from how a story you tell summarizes the history of your tribe. No essay can derive the biochemical story, or the cognitive one, or the social one. Here we point to a crucial moment where they all begin: not in the perception of detail, but in its selective destruction and lossy compression. The arguments of this essay suggest that averaging-out may have been the first source of memory, and thus self-reference, in the history of the universe. Perhaps that happened first in biochemistry; perhaps it had an even earlier start.

\section{Learning from the gaps}

The leaps the universe has made, from non-life to life, chemical reaction to mind, individualism to society, aimless to aim-ful depend in a basic way on how new features of the world---physical features, biological features, social features---become available for feedback and self-reference. If this essay is correct, then it is those self-referential features that, in creating predictive gaps, attract our curiosity. And it is those features that, at the same time, make the problems so hard. You might say we're constantly nerd-swiped by the origin gaps~\cite{xkcd}. 

Though I've focused on the primordial scene, the origin of memory, and located it in the coarse-graining of fundamental theories, I've also suggested that this coarse-graining process might be something worth attending to at later stages as well. This suggests an intriguing possibility: that there are more stages yet to come, new accelerations and ways for us to reflect upon ourselves, and (in doing so) to create new forms of life. It's natural, at this cultural moment, to look to the world of artificial intelligence and to ask what our machines will do for---or to---us. As we create machines with inconceivably greater powers to reflect, we may be setting in motion a process that will leave behind, for future millennia, a new origin problem to solve.

\clearpage

\section{Afterword: the Economics of Physical Law}

In the classical world, \emph{i.e.}, the world before we consider quantum effects, we describe the behavior of everything from planets to beachballs by talking about how they respond to the forces placed on them, and how they might create forces that others respond to in their turn. A basic feature of these laws, as far as we understand them, is that the only things we need to know are the positions of the particles, their velocities, and how their velocities change in time (their accelerations). A planet's gravitational field tugs on a beachball, causing it to descend; the effect of the planet on the ball can be summarized, without loss, by talking about how the ball accelerates in time. The position of the beachball (how far away it is from the center of the each) dictates its change in velocity with time (acceleration). In more complicated situations, such as magnetism, the acceleration of a particle might depend on its velocity as well as its position. But none of the laws we know of tell us, for example, give an independent role to (say) jerk, the change in acceleration with time.

A higher-derivative theory, by contrast, is one where facts about these more derived changes do have an independent causal power over the system's evolution. We can write them down, if we like, but when we examine the predictions they make, we find that they not only do not describe the world we know, but in important senses they can not describe anything that remotely looks like the world we know. We've known this since Mikhail Vasilevich Ostrogradsky published what is now called ``Ostrogradsky's Theorem'' in the journal of the Academy of St. Petersburg, in 1850. Richard Woodard has an excellent, technical introduction to Ostrogradsky's theorem on Scholarpedia~\cite{scholar}. 

Here, I'll try to give an intuitive introduction to the physics behind it, by drawing a parallel to economics. Physics and economics have long travelled in parallel. Students of the great physicist, and modern interpreter of thermodynamics, Josiah Willard Gibbs, went on to define 20th Century economics, at least for the Americans: Paul Samuelson was a direct descendent, who not only won the Nobel Prize in economics, but trained generations of policy-makers to come through his 1948 textbook Economics: An Introductory Analysis. But the parallels go back further in time, and in previous centuries yet it was not uncommon for someone today known for contributions to physics to have also been intrigued by, and often an originator of, basic concepts in economics. Let's go back to that tradition to see what happens.

Since the late 18th Century, physicists have defined theories by describing how particles move and arrange themselves in space so as to minimize a particular quantity. The quantity, called the \emph{Lagrangian}, attaches a value to every possible arrangement of particles in space and (crucially) to their velocities as well. Think of the Lagrangian as defining a cost that the particles pay for being in a certain place with a certain velocity. As these particles move through space, passing from one arrangement to another (or staying in the same configuration for a while) they run up a bill. Some configurations of course, are more costly than others; some are costly, but enable the particle to get to less costly configurations later.

To figure out how particles set in motion at one time (say, 9 am) will move and interact with each other, ending up in a new configuration at (say) 10 am, we consider all possible paths the particles might have taken to get from the arrangement at 9 am to the arrangement in question at 10 am. Each path runs up a bill, and the actual path that particle takes is that which runs up the smallest Lagrangian bill. Some paths are absurd, with particles stopping and starting at random, accelerating to vast speeds and just as quickly coming to a halt; these end up running up very large bills. The ones that rack up smaller bills are close to the paths the particles actually take, and the very smallest bill is the path the particles do take. (Hidden in here is the secret for generalizing to quantum mechanics: we now allow particles to stray from the optimal paths, penalizing them the further they stray, and indeed, all the arguments we're about to make remain valid for the quantum world.)

One of the puzzles of the Lagrangian formulation is that it's hard to think of this path selection as a causal story, in the usual sense: as a particle moves in space, it may be able to achieve a minimal bill by temporarily paying large costs ``in anticipation'' of reduced costs later in time---somewhat like a young financier living in New York City in his early twenties, only to move to Connecticut in mid-life. But in another sense, causality is preserved: only facts about the Lagrangian bill matter, and the only way to manipulate the particle is to alter, or add, terms in the Lagrangian. The particle doesn't really anticipate and think through the consequences of its behavior, although in the quantum mechanical formulation, you can think about it as trying a bunch of different paths in parallel universes.

In any case, ordinary Lagrangians only bill particles in terms of their positions and velocities: informally, it costs something to climb a hill, and it costs something to be fast. All paths can be defined in terms of their position and velocity alone; you can measure the implied acceleration, of course, at any point, but accelerations only matter to the extent to which they end up affecting the particle's velocity or position. Nothing else is billed: it comes for free, like the bread before the appetizer.

We can ask how much \emph{more} it costs to be fast, given a little boost in speed (\emph{i.e.}, when we take the derivative of the Lagrangian with respect to the velocity). If I'm already going 50 miles/hour, how much more does it cost to go to 51? If you're an economist, you can think of this as the \emph{marginal cost}: if I'm already making fifty thousand widgets, how much more does it cost to make the next?

Let's persist with the economic analogy. In general, the cost of an additional unit is different from the unit-by-unity costs you accumulated so far. Consider a drug company introducing a new drug to market: the cost to make the first tablet is enormous (the costs of researching, testing, and getting approval for the drug), while the cost to make the second tablet is much less: just leave the machine on for ten seconds longer. A similar example is the case of Amazon, who finds it cheaper to ship the millionth book than it did the first.\footnote{Making this article completely self-referential, the first book shipped from Amazon was by Douglas Hofstadter and the Fluid Analogies group at Indiana.} Conversely, consider asking a friend who works as a consultant for increasingly complex help: you might be able to get some brief advice for the cost of a cup of coffee, but if you want more than few minutes of thinking, you'll find that you'll start getting charged---indeed, more and more as the complexity of your problem becomes apparent. Another example is the declining marginal productivity of land: it's easier to feed the first fifty people in your village because you can farm the richest plots. As the population increases, you have to move to increasingly barren soil.\footnote{A more complicated relationship might obtain when hiring a taxi: the first mile is more expensive compared to the second mile, because you're usually charged a flat rate, or ``flag drop''. But if you try to get a taxi to take you from, say, Pittsburgh to Chicago, you'll find that you end up negotiating a much higher per-mile fee than you'd expect, since the driver won't be able to get a return fare. We'll focus here on the cases where every marginal cost specifies a unique unit amount---these also obtain for sensible physical laws, and our toy example of the drug firm. The technical term is that the Lagrangian is ``non-degenerate''.}

For any marginal cost, you can ask: how much am I saving (or losing) compared to the cost I \emph{would} have paid if that marginal cost applied at all levels? You can think of this as the foolish startup's price: if Amazon can find, pack, and ship a book for less than a dollar, my foolish friend reasons he can do the same with the books in his house. Or (if the curve goes the other way, with increments becoming more expensive) you can think of it as the freeloader's price; what happens when she notices that she only paid for an hour of advice, but actually got an hour and a half's worth once you count that free conversation over coffee. In the physical world, usually (but watch out!), it costs more to go from 50 to 51 miles/hour than it does from zero to one---so we can think of this as a freeloader's price; I got up to 51 miles/hour more cheaply on my Lagrange bill than I'd have expected given what I was charged for the last increment.

Now, remarkably, the freeloader's gain---the difference between the freeloader's price and the actual price---has an interpretation in terms of the underlying physics. It's the energy! 

Don't ask me \emph{why} it's the energy. It does all the sorts of things we want energy to do, like total your car if you get too much of it too quickly. If you stick in a theory that you can solve some other way, where you've previously been able to identify the quantity you think is the energy, it always comes out the same. But why it should pop out of a crazy argument about linear extrapolations of marginal costs, I can't say in any simple, efficient way. You might as well say that there's a hidden economic structure to the Universe that we didn't expect, and it turns out that energy is just some quantity derived from that more fundamental structure.

In any case,  and as long as the cost function doesn't change with time, you can prove that the total freeloader's gain in the system is constant---that's conservation of energy. Some people might get a few fewer hours of free consulting time, which lowers their freeloader's gain, but others, in turn, will get more. Something that we tend to think of as an essential quantity, neither created nor destroyed, ends up popping out of the Lagrangian formalism.

This story, about minimizing Lagrangian bills and freeloader's prices, may seem like an overly complicated way to talk about how particles move about in space. Famously, when the physicist Richard Feynman encountered it as an undergraduate during his physics education at MIT he rejected it entirely at first, coming up with increasingly ingenious ways to reason about physical systems using the standard set of Newton's laws. Why bother rephrasing the laws we already know in terms of a cost function? One answer is that it does provide a recipe for handling extremely complicated systems that you can't keep track of all at once. Even Feynman had, at some point, to switch over.

Another answer is that it provides a very general way to think about physical laws. All you need to do is specify that cost: a single equation, for example, can replace all three of Newton's laws of motion. Today, when inventing a new quantum field theory, all the physicist has to do is write down a Lagrangian. (Solving it, of course, is another problem altogether.) Most germane to our discussion, the Lagrangian formalism allows us to speculate on physical laws that have higher-derivative terms, providing a recipe book for how to interpret the equations in terms of physically real quantities like energy and momentum.

The effect of adding in these new terms is dramatic. This is because once you allow the Lagrangian to depend on more than just the velocity, but also the acceleration, you have multiple terms to consider. The Lagrangian depends on not just the cost of going from 50 miles/hour to 51, but also the cost of going from (say) zero acceleration to 1 mile/hour/second. 

To continue the economic analogy, there are more goods to produce, and by adjusting the mixture of goods one produces, unexpected cost savings become possible. Following the standard recipes shows that it's now possible to find economies of scale: as one speeds up, for example, it becomes easier and easier to speed up more. We move from the village farm to the Amazon warehouse. These unexpected gains correspond to negative energies: rather than costing energy to get there, you actually release a little.

This sounds harmless at first. But there are still positive energy paths. And, although energy can neither be created nor destroyed, we're now in a system where particles can charge arbitrarily large gains to other particles, as long as they can match foolish startup prices to freeloader's gains. A car can accelerate to an arbitrarily high speed and high energy, from nothing, as long as it can find another car who can produce negative energy to keep the totals constant. You might say we've invented debt, and given the particles, like Lehman Brothers, no constraints on how much they can leverage. 

\section{A conversation with John Bova, Dresden Craig, and Paul Livingston}

The ``Undecidables''~\cite{undec} met on 5 July 2017 in Santa Fe to discuss a draft of this essay. John Bova, Dresden Craig, Paul Livingston and I participated, in a meeting that also touched on papers by John~\cite{bova} and Simon Saunders~\cite{simon}. In the following weeks, the group proposed a series of questions based on that conversation, which I've attempted to answer here.

\vspace{0.5cm}

\noindent
{\bf Dresden Craig}. In the abstract for this paper, you write that ``the universe need not have made self-reference possible.'' Could you say a little more precisely what you mean by that? Do you mean ``the universe'' there to indicate a universe with the same fundamental laws of physics as our own, or are you also thinking about other possible universes? (If the latter, ``possible'' in what sense?) In either case, can we say anything meaningful about what a universe in which self-reference was impossible would be like?

{\bf SD}. My main concern here was with universes that had laws, physical laws, that differed from ours but nonetheless could be expressed in the formalisms and mathematics we have to hand in our own. The story of modern physics is in part the story of how we've come to learn that---if we value mathematical coherence---this space is much more restricted than we used to think, and that's a lot of fun. There are certainly strange universes well beyond that, that philosophers have considered, but to include all of those as well is a bit cheating. It's trivial to consider, say, a universe consisting only of a perfect sphere, hanging unaided in an infinite space. There's not a lot going on there.

What you do need for self-reference are structures sufficiently densely-interlocked that they support an effectively unbounded memory. The natural way for this to happen, in our universe, is through the sticking together of stuff in increasingly complex ways, with memories being spread out over increasingly large distances. Consider the leap from a protocell, with a simple membrane to separate in from out, to the human brain. You don't even need a Lagrangian in your fundamental theory for that to happen: all sorts of crazy theories could do that, including whatever M-Theory turns out to be. But, conversely, it's possible to write down universes where this stuff would not happen: \emph{e.g.}, a billiard-ball universe without gravity.

It's worth noting that there are many ways to get this interlocking, some of which would seem very strange to us. Imagine, for example, things clustering together not in space, but in ``momentum space''---nearby by virtue of having similar values of momentum, like cars grouped not by where they were on the highway, but by their speeds. You could imagine physical laws tuned in such a way that particles with particular momenta were able to preferentially interact even if widely separated (or delocalized) in space. Memory could emerge.

\vspace{0.5cm}

\noindent {\bf Paul Livingston}. My first question is about physics, memory, and time.  As you note in your paper, fundamental physics doesn't appear to allow for memory to be a basic (or even a real) phenomenon: since the time parameter in statements of physical law is always just given as a simple, single value increasing over time, there doesn't seem any warrant for introducing as physically real any operations of comparison between states of systems at distinct times.  Things seem (at least at first) rather different from the perspective of computer science, where of course we constantly appeal to data being stored ``in memory'', and even Turing's basic architecture essentially includes an (ideally infinitely extended) symbolic memory.  

At this level, we have memory in the sense of the ability to store syntactic symbols, and for the machine's functioning at one time to depend on what has previously been stored; but we don't yet seem to have a basis for at least some of the further emergent phenomena you discuss in your paper---for instance meaningful experience---until these symbols and comparisons are in fact ``interpreted" by some kind of conscious subject or agent. If this is right, it would seem to make the presence of such an agent (who lives in experienced time) essential for these phenomena themselves, as if in an important sense it is us who are constituting or making up time (beyond just the single, linear time-parameter of basic physics).

In the history of philosophy, there's a long legacy of arguments that say that time is not basically real, or is illusory at the basic level of reality and is rather constituted by us as human subjects.  These arguments perhaps begin with Zeno, who held, for example, that an arrow in flight cannot really be moving, since at each discrete moment (each discrete value of t) it is at rest.  Others such as Kant and Husserl have seen time and the meaningful phenomena of change, motion, and causality as imposed by the form of our minds or our understanding upon the world, while still others (such as Bergson) have tried to re-introduce ``memory" into matter by thinking of universal time, including physical time, as constituted in part by a kind of universal cosmic ``evolution" toward progressively higher forms.  

My question then is whether and how the dynamics of self-reference can allow us to see time and memory as really ``there" (at the basic physical and/or computational levels) and not just constituted or produced by us.  And can we maintain this kind of realism about time and memory without thinking that all of the progressive developments that you've invoked (life from non-life, mind from non-mind, and society from individuals) were ``pre-inscribed," that is, already built in to the basic physics of the universe, somehow?

{\bf SD}. Modulo some minor translations between our two languages, I'd agree with your account of time plays here, and your extension to the experience of conscious agents. It certainly feels like it's impossible to experience the passage of time without some kind of reference to past and future, and such comparisons become impossible for an agent without the memory to do so, or a subject without access to those memories.

Once we phrase it this way, I come down squarely on the side you attribute to Kant and Husserl. We can knock off Zeno right away, if with a cannonball from the 19th Century. The Danish physicist Hans Christian \O rsted showed that an electrical current---meaning the flow of electrons through a wire---could move the needle on a compass sitting a little distance away by inducing a magnetic field. We now believe that effect doesn't depend in any fundamental way on the inhomogeneity of the flow. If you pushed a perfectly uniform, charged rod, infinitely long, past the compass, a similar thing would happen, but now the set of discrete moments for that moving rod are identical to the case where it's stationary. The only thing you're doing to  is altering the velocity, and yet that difference causes a needle to twitch. So velocity is real, since anything that plays a causal role has to be real, and since position is also real and velocity is the derivative of position with time, there is some non-trivial sense in which time is a physical thing that just exists, does something in the world. Position and velocity are two legs of a tripod, and the whole thing can only stand with time.

But at the same time, we do want to say that what \emph{we} mean by time---the feeling of time flowing up towards a deadline, say, or away into the past after a parting from someone you love---is more than just a component of physical law. It essentially involves the awareness of change, internal or external, the ability to compare one moment to the next, and to consider its meaning. We can't understand the world, make sense of it, without grasping these more complex objects---objects that, by Ostrogradsky's theorem, are banned from being fundamental constituents of our world. They emerge. I'd be happy to claim Bergson as a fellow-traveller here, since I want to say that memory and the possibility of self-reference can emerge prior to a subject; they can become available, we can have them to hand.

Now, I can't quite parse the Subject from the computation. I wish I could. But we can split the physical and computational apart, and say that you need the computational bit to gain awareness of time, and that computational bit can't be ``baked in'' like the time that \O rsted discovered.

\vspace{0.5cm}

\noindent {\bf DC}. You write that ``the fundamental laws [of physics] are forgetfulÓ, referring to how third (or higher) time derivatives cannot correspond to physical realities; then, you write that ``coarse graining gives the possibility of memoryÓ and later you ``point to a crucial moment where they [\emph{i.e.}, each new level of complexity] all begin: not in the perception of detail, but in its selective destruction and lossy compression.'' Is this a fair paraphrase? Memory is made possible by a kind of coarse graining, and coarse graining is a throwing away of details; memory is, therefore, a particular form of forgetting. If so, then can we say the inability to derive certain key thresholds of complexity from fundamental physical laws is tantamount to an inability of physics to predict what about itself is forgettable? Can we generalize to say that any theory which works for a given level will be unable to predict which coarse grainings of itself will make sense?

{\bf SD}. This is a lovely question---and a lovely suggestion. One has a feeling that (for example) humans are constantly surprised by the coarse grainings society places on them: categories of race and sex, class, and so forth. ``I wouldn't have ever imagined you'd do that to us,'' you can hear people say at critical points in history, referring not to a particular other person, but rather to a system that they find themselves caught up in. The understanding we have of our own inner lives can't work out what society will do with collections of them. 

It may well be the case that we recognize the emergence of new things only in retrospect, when we have the examples before us; and even then, only imperfectly. And of course we can disagree about which coarse-grainings are appropriate. Historians generate multiple accounts of the same events, which you can think of as incompatible coarse-grainings, and they find the clashes between these accounts to be sources of fertile discussion, rather than signs that something is seriously wrong with the project.

\vspace{0.5cm}

\noindent {\bf PM}. Throughout your paper, I was interested in the way that issues about how we describe the phenomena theoretically interact with issues about how these phenomena actually are in themselves.  For example, one can argue that a computer's memory register is only actually its memory as described from a certain (functional) perspective: from another perspective, it is just a physical configuration of matter. We might distinguish here between the first and second of Marr's levels of analysis for systems.\footnote{Marr, a neuroscientist, described three levels for the analysis of human visual processing: the ``computational'' (what purpose is the system achieving---recognizing faces, say), ``algorithmic'' (how does the system break that task into subtasks that fit together? can you write down the psuedocode of that process?), and ``implementation'' (how the brain actually gets things done, with Potassium ions and depolarization waves)~\cite{marr1982computational}.} If we draw the distinctions this way, it seems that a system can only be ``self-referential" (if it is) as described in a certain way: that is, for us to see it as self-referential we must describe it from a perspective that portrays some of its physical changes and quantities as ``references" to things, and also draws some line around just that system as "itself."  

This might seem to suggest that while the idea of self-reference can change the way we view some of these systems, it doesn't play a role in the basic, physical-level behavior of the system itself.  Yet all of the phenomena you describe (life, mind, and society, etc.) do certainly---once they are there---make a difference to the actual physical behavior of the relevant systems: for example, given a functioning society, matter will be moved to places it would not be if there were just the individuals acting without any conception of the larger whole.  How should we understand, then, the actual reality of these features of self-reference and self-organization, at the most basic physical level?  Or does acknowledging them require us to hold that a total explanation of the world written only in the vocabulary of the basic physical level (without terminology such as ``reference" and ``self-") could never capture all that goes on there? 

{\bf SD}. I think the Marr account is the correct one. We can get these terrific, efficient descriptions of how a pile of mail got from one side of the Atlantic to the other by referring to emergent properties. We could re-write everything in a more fundamental language, but why? We do so much better if we don't. You might say that we know the reality of reference and self from the epistemic resistance we encounter when we try to get rid of them.

\vspace{0.5cm}

\noindent {\bf John Bova}. Is it possible to take us a little farther into the discussion of renormalization? And how do the considerations about renormalization relate to the simpler reasons that we might expect the fundamental laws of medium-sized nature (at least) to work on the order of second derivatives? For instance, it seems as though there ought to be a connection to how a conservative field can be understood as throwing away path-dependent information.

{\bf SD}. There are many different ways things can become ignorable. When we write down a theory, we encounter quantities that are ``truly'' irrelevant: for example, in classical electrodynamics, the zero point of the electric potential. You can move those quantities around as much as you like, and the predictions of the model don't change. We call the gauge invariances. They were never there in the real world in the first place, but were rather ghosts born of our limitations. The structures in the world that we want to describe don't map perfectly onto the mathematical objects we know how to manipulate. There's an excess, though sometimes we find better notation that kills them off once and for all.

In other cases, we have facts about the world that are real, but causally irrelevant. The motion of a particle in a conservative field, as you note, can be both predicted and explained without reference to the entire path it took. It's not that the path doesn't have some kind of reality. You and I can watch the Space Shuttle launch, in a way that we can't, for example, observe the zero point of the electric potential. It's just that the path is irrelevant from the point of view of the physics of the phenomenon itself. A similar thing happens for jerk, in fundamental physics: it's not that you can't take the third derivative of position, it's just that it doesn't matter.

Renormalization is a third, distinct, way of ignoring things. In this case, you actually end up ignoring things that matter! You pick and choose carefully so that what you're ignoring is (for example) the least-damaging, least problematic stuff. If you're interested in building a road, you don't need to have the positions of the quarks, but they're really there and they matter causally to what's going to happen to your road. Renormalization does a huge number of things in physics, beyond just the production of efficient descriptions and (as we use it in this paper) the discovery of higher-order, non-local interactions. One of the first reasons we created it was to handle some problems with a theory that we thought was fundamental (quantum electrodynamics, QED) but had all sorts of problems when we tried to treat it that way. In the end, we had to create the Higgs boson---but we were able to do QED calculations before we knew what the Higgs was, because we could treat QED as a theory that came out of some mysterious mystery theory deeper down.

I'd be remiss if I didn't mention the Santa Fe Institute MOOC we did on this~\cite{renorm}; one of the papers on renormalization in complex, computational systems that we talk about there comes from Israeli and Goldenfeld~\cite{israeli}.

\vspace{0.5cm}
\noindent {\bf JB}. Will a human sorter faced with the Shakespearean task produce a locally sharp but globally indefinable shape, or will they produce a hopelessly vague shape even at the level of local decisions? What does the answer tell us about how to apply theorems on formal systems to human intentional states and acts without conflating or merely analogizing the two? What does it tell us about where intentional states are located in conceptual space relative to consistency and completeness? 

{\bf SD}. You're asking a crucial question. Paul asks it from a slightly different angle, making a distinction between the computation and the subject, while you distinguish formal system from human reason. How do we, as subjects, differ from our computational states? Or (a more restricted question) where are \emph{we} located, if we are at all, on the Marrian levels?  The purpose of this essay is to give a story about how a certain set of objects (memory, self-reference) that are necessary (but not sufficient) for meaning, come to be, and it's tempting to punt the rest to biology, tying the intentionality and meaning of human states to evolution. But let's try a bit harder before we punt.

What fails when we think that formal systems are identical to human states of mind, thoughts a subject thinks? An obvious answer is the latter are subject to continuous revision. I may hold in mind the same object, person, thought, at two different times, and have them mean, or imply, completely different things. We develop, over time, in ways that I think can't be simply reduced down to computational talk about changing variables, or even LISP-like metaprogramming stories. We're not substituting, we're returning.

Let answer this synthesis of your question and Paul's with an answer suggested by Dresden. What if we take seriously the revisability of our our mind's meanings, the ways in which we can return to the same states, the same thoughts, and have them mean completely different things to us? Here let me include our emotional responses, our feelings, as well as our more conceptual, intellectual states. 

We could say, well, it's just the valence that's changed, the thing is the same but it's stuck in a different place of our affective network. The context is different. But I think it's more than this: there's something unsatisfying about postulating a world of atomic thoughts that we combine together like Lego bricks, even if we made those bricks ourselves. At the very least, we experience a shift in emotional gestalt, where we might in some way be able to say the object is identical, but not in any way that actually matters. We feel love for a person, but where this was originally a pleasure, it has now become a pain, because that person is gone from us in some way. And it becomes a pleasure or a pain in a very different way from how we perceive food differently depending on how hungry we are.

Put these pieces together and we have a problem. How can we have a simple object (love, say---but if you like, one of the components of love) that persists in time (the love now is the love then), but also changes in some essential quantities (it was a pleasure, now it's bittersweet)? One answer is if the coarse-graining of the fundamental constituents is changing. At the level you're experiencing love, it's a simple object, whose complexity is hidden from you by the coarse-graining process. Events shock you, knock you about, and that coarse-graining has shifted in scale. The nature of that shift allows you to track the object over time, as its properties change, or at least to rediscover it after a shock. Perhaps it zooms out, so that the difficult parts of that love become invisible; perhaps it zooms in a bit, magnifying feelings you didn't know you had; most likely, some combination of the two. The shapes of our minds, and how they map to some computational or formal-language account, are not just vague. They're shifting. That makes formal systems inadequate not just as descriptions of our actual function (we already knew we weren't perfect reasoners), but also as normative accounts of our emotional lives.

\clearpage
%\bibliographystyle{unsrt}
%\bibliography{fqxi}

\begin{thebibliography}{10}

\bibitem{turkle2011life}
Sherry Turkle.
\newblock {\em Life on the Screen}.
\newblock Simon and Schuster, New York, NY, USA, 2011.

\bibitem{walker2013algorithmic}
Sara~Imari Walker and Paul~CW Davies.
\newblock The algorithmic origins of life.
\newblock {\em Journal of the Royal Society Interface}, 10(79):20120869, 2013.

\bibitem{kauffman2016humanity}
Stuart~A Kauffman.
\newblock {\em Humanity in a creative universe}.
\newblock Oxford University Press, Oxford, United Kingdom, 2016.

\bibitem{hobson2015social}
Elizabeth~A Hobson and Simon DeDeo.
\newblock Social feedback and the emergence of rank in animal society.
\newblock {\em PLoS Computational Biology}, 11(9):e1004411, 2015.

\bibitem{major}
Simon DeDeo.
\newblock Major transitions in political order.
\newblock In S.I. Walker, P.C.W. Davies, and G.F.R. Ellis, editors, {\em From
  Matter to Life: Information and Causality}. Cambridge University Press,
  Cambridge, United Kingdom, 2017.
\newblock Available at \url{https://arxiv.org/abs/1512.03419}.

\bibitem{aaronson2014quantifying}
Scott Aaronson, Sean~M Carroll, and Lauren Ouellette.
\newblock Quantifying the rise and fall of complexity in closed systems: The
  coffee automaton.
\newblock {\em arXiv preprint cond-mat.stat-mech}, 2014.
\newblock Available at \url{https://arxiv.org/abs/1405.6903}.

\bibitem{aaronson2013philosophers}
Scott Aaronson.
\newblock Why philosophers should care about computational complexity.
\newblock In {\em Computability: Turing, G{\"o}del, Church, and Beyond}, pages
  261--328. MIT Press, Cambridge, MA, USA, 2013.

\bibitem{chomsky2014aspects}
Noam Chomsky.
\newblock {\em Aspects of the Theory of Syntax}.
\newblock MIT Press, Cambridge, MA, USA, 2014 [1964].

\bibitem{smith1997major}
John~Maynard Smith and E\"{o}rs Sz\'{a}thmary.
\newblock {\em The Major Transitions in Evolution}.
\newblock Oxford University Press, 1997.

\bibitem{fukuyama2011origins}
F.~Fukuyama.
\newblock {\em The Origins of Political Order: From Prehuman Times to the
  French Revolution}.
\newblock Farrar, Straus and Giroux, 2011.

\bibitem{smail2007deep}
Daniel~Lord Smail.
\newblock {\em On deep history and the brain}.
\newblock University of California Press, 2007.

\bibitem{bellah2012axial_merlin}
Merlin Donald.
\newblock An evolutionary approach to culture.
\newblock In Robert~N Bellah and Hans Joas, editors, {\em {The Axial Age and
  its consequences}}. Harvard University Press, 2012.

\bibitem{graeber2014debt}
David Graeber.
\newblock {\em Debt: the first 5,000 years}.
\newblock Melville House, New York, NY, USA, 2014.
\newblock Updated and expanded.

\bibitem{woodard2015theorem}
Richard~P. Woodard.
\newblock The theorem of {O}strogradsky.
\newblock {\em arXiv:1506.02210}, 2015.
\newblock Available at \url{https://arxiv.org/abs/1506.02210}.

\bibitem{dedeo2008stable}
Simon DeDeo and Dimitrios Psaltis.
\newblock Stable, accelerating universes in modified-gravity theories.
\newblock {\em Physical Review D}, 78(6):064013, 2008.

\bibitem{cooney2009gravity}
Alan Cooney, Simon DeDeo, and Dimitrios Psaltis.
\newblock Gravity with perturbative constraints: Dark energy without new
  degrees of freedom.
\newblock {\em Physical Review D}, 79(4):044033, 2009.

\bibitem{dedeo2011effective}
Simon DeDeo.
\newblock Effective theories for circuits and automata.
\newblock {\em Chaos: An Interdisciplinary Journal of Nonlinear Science},
  21(3):037106, 2011.

\bibitem{flack2012multiple}
Jessica~C Flack.
\newblock Multiple time-scales and the developmental dynamics of social
  systems.
\newblock {\em Philosophical Transactions of the Royal Society B: Biological
  Sciences}, 367(1597):1802--1810, 2012.

\bibitem{Flack20160338}
Jessica~C. Flack.
\newblock Coarse-graining as a downward causation mechanism.
\newblock {\em Philosophical Transactions of the Royal Society of London A:
  Mathematical, Physical and Engineering Sciences}, 375(2109), 2017.

\bibitem{xkcd}
Randall Munroe.
\newblock Nerd-swiping.
\newblock Available at \url{http://xkcd.com/356}.

\bibitem{scholar}
R.~P Woodard.
\newblock {O}strogradsky's theorem on {H}amiltonian instability.
\newblock {\em Scholarpedia}, 10(8):32243, 2015.
\newblock revision \#151184.

\bibitem{undec}
John Bova, Dresden Craig, Simon DeDeo, and Paul Livingston.
\newblock The {U}ndecidables, 2013---.
\newblock Archive available at
  \url{http://tuvalu.santafe.edu/~simon/undecidables.txt}.

\bibitem{bova}
John Bova.
\newblock Groups as eide? toward a {P}latonic response to {M}etaphysics {M} on
  unity, structure, and number, 2017.
\newblock Unpublished manuscript.

\bibitem{simon}
Simon Saunders.
\newblock Physics and leibniz's principles.
\newblock In K.~Brading and E.~Castellani, editors, {\em Symmetries in Physics:
  Philosophical Reflections}. Cambridge University Press, Cambridge, United
  Kingdom, 2003.

\bibitem{marr1982computational}
David Marr.
\newblock {\em A computational investigation into the human representation and
  processing of visual information}.
\newblock Freeman and Company, San Francisco, CA, USA, 1982.

\bibitem{renorm}
Simon DeDeo.
\newblock Introduction to renormalization, 2017.
\newblock Complexity Explorer MOOC, Santa Fe Institute.
  \url{http://renorm.complexityexplorer.org}.

\bibitem{israeli}
Navot Israeli and Nigel Goldenfeld.
\newblock Coarse-graining of cellular automata, emergence, and the
  predictability of complex systems.
\newblock {\em Phys. Rev. E}, 73:026203, Feb 2006.

\end{thebibliography}

\end{document}